\documentclass[twocolumn,showpacs,preprintnumbers,amsmath,amssymb]{revtex4}
\usepackage{graphicx}

\begin{document}
\def \be{\begin{equation}}
\def \ee{\end{equation}}
\def \bea{\begin{eqnarray}}
\def \eea{\end{eqnarray}}
\def \half{{1\over 2}}
\def \etal{{\it et~ al~}}
\def \cH{{\cal{H}}}
\def \cM{{\cal{M}}}
\def \cN{{\cal{N}}}
\def \bS{{\bf S}}
\def \bL{{\bf L}}
\def \e{{\epsilon}}
\def \a{{\alpha}}
\def \t{{\theta}}
\def \b{{\beta}}
\def \g{{\gamma}}
\def \D{{\Delta}}
\def \d{{\delta}}
\def \w{{\omega}}
\def \s{{\sigma}}
\def \nd{{^{\vphantom{\dagger}}}}
\def \yd{^\dagger}
\def \ket#1{{\,|\,#1\,\rangle\,}}
\def \bra#1{{\,\langle\,#1\,|\,}}
\def \braket#1#2{{\,\langle\,#1\,|\,#2\,\rangle\,}}
\def \expect#1#2#3{{\,\langle\,#1\,|\,#2\,|\,#3\,\rangle\,}}
\def \rl#1#2{{\,\langle\,#1\,#2\,\rangle\,}}
\draft

\title{ Singlet Excitations  in Pyrochlore:  A  Study of  Quantum Frustration}
 \author{Erez Berg, Ehud
Altman and Assa Auerbach}
 \affiliation{ Department of
Physics, Technion, Haifa 32000, Israel.} \date{\today}
\begin{abstract}
We apply the Contractor Renormalization (CORE) method to
the  spin half Heisenberg
antiferromagnet on the  frustrated checkerboard and  pyrochlore lattices.
Their ground states are spin-gapped singlets which break
lattice symmetry.  Their effective Hamiltonians  describe fluctuations of
orthogonal singlet  pairs on  tetrahedral blocks, at an emergent low energy
scale.  We discuss  low temperature thermodynamics and new
interpretations of finite size numerical data. We argue that our results are
common to many models of quantum frustration.
 \end{abstract} \maketitle

\begin{figure}[h]    \centering
  \includegraphics[width=9cm]{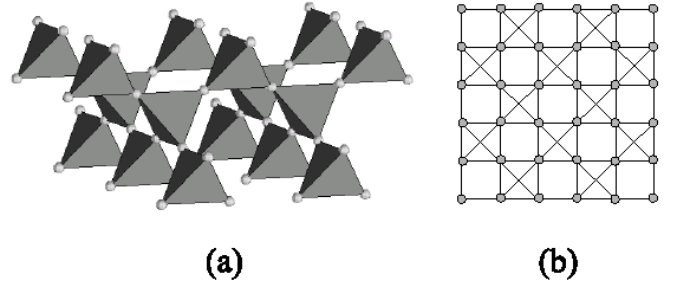}
  \caption{The Pyrochlore (a) and Checkerboard (b) lattices.}
  \label{fig:lat}
\end{figure}

Frustration in classical spin models often  leads  to a  complex  energy
landscape. Certain models, such as the Heisenberg antiferromagnet
on the pyrochlore lattice,  has an extensively degenerate ground state
manifold. This model given by
\be
H=J\sum_{\langle
ij\rangle}\bS_i\cdot\bS_j
\label{eq:heisenberg}
\ee
has  spins
 $\bS_i$ sitting on corner sharing tetrahedral units (see Fig.
\ref{fig:lat}(a)).

In the  {\em semiclassical}
approximation\cite{villain,henley,moessner-chalker},
a large  degeneracy survives the quantum fluctuations, and
thus resists  ground
state selection by
 the "order from disorder" mechanism.

A pressing  open question  is what happens in the
{\em  strong}  quantum limit, e.g.  the spin half case? 
Series expansions\cite{canals-lacroix} suggest rapid decay of spin correlations.
Does this indicate the formation of a translationally invariant spin liquid or lattice symmetry
breaking valence bond solid?
In the quantum case, is there an emerging
low energy scale,  in lieu of the
classical ground state degeneracy?

 The
 purpose of this Letter is to derive
the  low energy effective Hamiltonian
  starting from the Heisenberg
model. As a warm-up  to the  pyrochlore lattice
(Fig. \ref{fig:lat}(a)) , we  treat its  two dimensional (toy model) reduction,
the Checkerboard lattice (Fig. \ref{fig:lat}(b)). The Checkerboard  has
recently received significant theoretical
attention\cite{palmer,fouet,lieb,sondhi,canals}.

Our approach is  the Contractor Renormalization (CORE) method\cite{core}.
The CORE  is a real-space discrete renormalization transformation invented by
Morningstar and Weinstein.  It maps a  lattice Hamiltonian  to an
effective Hamiltonian with the same
low energy spectrum. The CORE method computes the  effective interactions at
all ranges using exact diagonalizations of finite connected clusters.
Truncation of   interactions beyond a finite
range  is an approximation whose error can be estimated numerically from the
next higher range terms. CORE  has  been  successfully applied to describe the
spectra of  Heisenberg models on chains and ladders\cite{core,PS}. Recently,
it was applied  to  the square lattice  Hubbard model  to derive the Plaquette
Boson-Fermion Model for cuprate superconductors\cite{PBFM}.
We refer the reader to previous reviews\cite{core,PBFM}  for the mathematical
background and technical details.

For each of the Hamiltonians  at hand, we  define   local
operators  from the lowest eigenstates of  the elementary
clusters, e.g. a tetrahedral unit in the Pyrochlore lattice. We
shall compute the effective interactions  by CORE up to four
clusters range, and estimate the truncation error.

\begin{figure}[t]
  \centering
  \includegraphics[width=7cm]{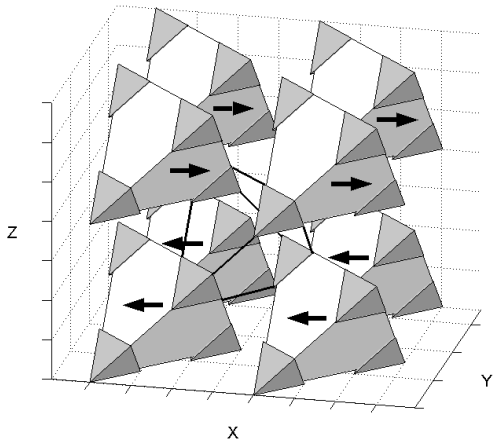}
  \caption{The pyrochlore viewed as a cubic lattice of
  supertetrahedra. The arrows show the direction of the
  supertetrahedra's pseudospin in the MF ground state.}
  \label{fig:pyrogs}
\end{figure}

Our  key results  are the following:  {\em For the Checkerboard,}   we confirm
the conclusions of refs \cite{sondhi,canals}, derived by other means, that
the ground state is closely approximated by the product of
uncrossed plaquette singlets.
The effective pseudospin Hamiltonian allows us  to  interpret  the numerical
spectrum of  low lying singlets\cite{fouet,palmer} in terms of   Ising domain
walls. This yields the  number of
singlets as a function of lattice size:   a numerically testable prediction. In
addition we expect  a branch of weakly dispersive triplets at an
energy scale slightly  below the Heisenberg exchange.
{\em For the  Pyrochlore,} we apply two CORE steps  to arrive at an
Ising-like model of local singlets on larger  super-tetrahedra blocks which
form   a cubic superlattice.  At mean field level, we obtain a  singlet
ground state which breaks lattice symmetry as depicted in
Fig.\ref{fig:pyrogs}.
Here too, the effective Hamiltonian describes
Ising-like domain walls.   We shall conclude that lattice symmetry breaking
and local singlet  excitations
are general features of highly frustrated quantum  antiferromagnets.

{\em The Checkerboard.}  The lattice depicted in Fig.  \ref{fig:lat}(b),
contains crossed plaquettes (which are equivalent to  three
dimensional tetrahedra)
 connected by uncrossed plaquettes.
The first step is to choose elementary clusters which cover the lattice. Two
different options for plaquettes are the  {\em crossed}  and the  {\em
uncrossed
 plaquettes}. While the two clustering choices  may appear to yield
different
 ground states and excitations, we shall see
 that they are in fact
consistent with each other, and yield complementary information.

{\em Clustering with uncrossed plaquettes}. From the spectrum of a
single uncrossed plaquette, we retain the singlet ground state as
a vacuum state $\ket{\Omega}_i$ and the lowest triplet
as a singly  occupied boson  state $t\yd_{\a i}\ket{\Omega}_i$ .
$\a=x,y,z$ is a cartesian index of the
triplet.

The  effective Hamiltonian in the uncrossed plaquettes  basis  is
 (in units of
$J=1$):
  \bea
\cH_{eff}=&&\e_t'\sum_it_{\a i}\yd t_{\a i}\nd
+K\sum_\rl{i}{j}\bL_i\cdot\bL_j \nonumber
\\
&+&\sum_{\rl{i}{j} \a \b}\{-\D t\yd_{\a i}t\yd_{\a j}t\nd_{\b
i}t\nd_{\b j} + bt\yd_{\a i}t\nd_{\a i}t\yd_{\b j}t\nd_{\b j}\},
\nonumber \\ \label{eq:Heff-bose}
\eea
 where   $\bL_j=\sum_{\a\b}t\yd_{\a j}
\vec{\cal{L}}_{\a\b}t\nd_{\b j}$ and $\vec{\cal{L}}_{\a\b}$ are
$3\times3$ spin-1 matrices in a cartesian basis.

The parameters calculated by CORE  upto  range 2 are:  $\e_t' =
0.5940$, $K=0.2985$,  $\D=0.1656$ and $b=0.0776$. The truncation
errors from up to range-4 are less than $2\%$, and will be
ignored\cite{long}.

Note that  $\cH_{eff}$ in  (\ref{eq:Heff-bose})  commutes with the
number of triplets since it has no  anomalous pair creation
terms, as appear e.g. for  the square lattice \cite{PBFM}.  Thus,
at this level of truncation, the plaquette vacua product \be
\vert \Psi_0 \rangle = \prod_i \vert \Omega_i \rangle, \ee {\em
is an exact ground state of the effective Hamiltonian}
(\ref{eq:Heff-bose})\cite{comm-MG}.
 This result agrees with  Moessner \etal
\cite{sondhi}, who argued for a plaquettized  singlet ground state based on an
effective
 quantum dimer model.

We are also able to obtain the triplet (spin)  gap for
Since $t\yd_{\a,i}\vert 0 \rangle$ is an approximate eigenstate of
(\ref{eq:Heff-bose}), its
energy (spin gap) can be read from
$\e_t'=0.5940$.   This  compares well with
the value of 0.6-0.7 estimated by exact diagonalizations of
finite systems \cite{fouet}.
 We have found very weak hopping terms (of
magnitude $0.01J$)  due to  CORE interactions of  range four, which will
 give the triplets a weak dispersion in the full lattice.

\begin{figure}[b]
  \centering
  \includegraphics[width=6cm]{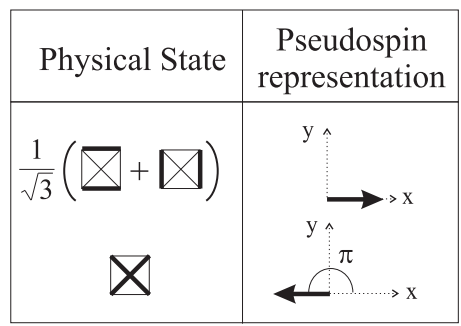}
  \caption{The two singlet ground states of the uncrossed plaquette and their
pseudospin representation. In the physical state thick lines denote valence
bond singlets.}
  \label{fig:spec-tet}
\end{figure}

{\em Clustering with crossed plaquettes}. The isolated   crossed
plaquette has two fold  degenerate singlet ground states, which
we can represent by a pseudospin-$\half$ doublet (see Fig. \ref{fig:spec-tet}).

The quantization axis for the pseudospin operators is chosen as
in Ref. \cite{tsunetsugu}, with the $+z (-z)$ directions
representing states with positive (negative) chirality. The
planar angles $0,\pi/3,2\pi/3$,   represent the three (non
orthogonal) dimer configurations of the tetrahedron. The states
with their pseudospin polarized in the $+x$ and $-x$ direction
are shown in Fig. \ref{fig:spec-tet}.

 The effective Hamiltonian  in the crossed plaquettes basis is
an  Ising-like model
 \be H_{eff}=-J'\sum_{\langle
ij\rangle}(\bS_i\cdot{\bf e}_{ij}) (\bS_j\cdot{\bf e}_{ij})
-h\sum_i S_i^x. \label{eq:Heff-crossed} \ee
where ${\bf e}_{ij}$ are directors on the $x-y$ plane pointing 
$\pi/3$ ($-\pi/3$) away from the $x$ axis for horizontal (vertical) bonds. 
At range-2 we obtain,
$J'=0.527665$, and $, h=0.118084 $
Corrections from range-3 and range-4
CORE were computed\cite{long}, and found to be unimportant for the symmetry
and correlations of the low excitations.

\begin{figure}[h]
  \centering
  \includegraphics[width=5cm]{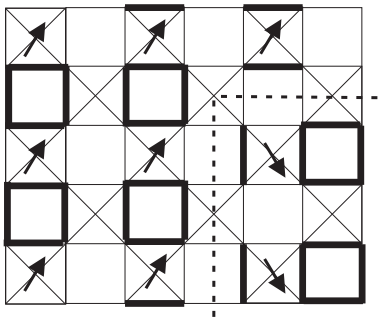}
  \caption{A singlet excitation of the checkerboard.
  This state is a
  domain wall (dashed line) between domains of the two plaquettized
  ground states. The corresponding pseudospin directions in the crossed
  plaquette approach are depicted by arrows.}
  \label{fig:domain-wall}
\end{figure}

We can solve (\ref{eq:Heff-crossed}) in mean field theory. The
energy exhibits two minima, where the pseudospins describe
vertical or horizontal dimers. These states correspond
respectively to projections of the two equivalent plaquette
ground states onto the truncated Hilbert space of the crossed
plaquettes. Although the ground  state energy is not well
converged at range-2 CORE,  (\ref{eq:Heff-crossed})  treats the
two symmetry breaking ground states in an unbiased fashion. It
therefore describes  low energy singlet excitations which are
pseudospin-flips  or Ising domain walls between ground states
(see Fig. \ref{fig:domain-wall}).

Thus we see that  the two methods elucidate
 complimentary aspects of the
Checkerboard. The approach using
 uncrossed plaquettes gives a very
accurate description of one
 ground state, which serves as its vacuum, but
where the other ground state is a  multi-magnon bound state. In
contrast, the crossed plaquettes describes correctly the low
energy singlet excitations by an effective Ising model
 $J^{Ising}S^z S^z$ with a coupling constant
 $J^{Ising}=9/8 J'$
renormalized by the quantum
 fluctuations. The lowest excitations are gapped
spin flips of  energy  $J^{Ising}$. They gain  a weak dispersion due to
the effective $S^+S^-$  couplings.

The number of singlet states
is expected to grow as   {\it power laws} with
the size of the system $N$. For example,the lowest lying
single spin-flips grow  as $N$, while higher
spin flip pairs  grow   as $N^2$ and so on.
An Ising-like phase transition is expected between the  broken and
unbroken  translational symmetry phases at a temperature scale of  $T_c\simeq
J^{Ising}$, with an associated logarithmic divergence of the
heat capacity at $T_c$.

{\em The Pyrochlore, } depicted in Fig. \ref{fig:lat}(a), is a
three dimensional network of corner sharing tetrahedra.
 Like the checkerboard, it has a
macroscopically degenerate (exponential in lattice size)
classical ground state manifold. For the  quantum S=1/2 case,
local pseudospins can be defined by the degenerate singlets of
disjoint tetrahedra. These cover all sites of the Pyrochlore and
form an FCC superlattice.
The effective hamiltonian on this FCC lattice was calculated by
CORE. The first non-trivial inter-teterahedra coupling are
obtained  at range three connected tetrahedra, which yield

\begin{eqnarray}
\label{eq:pyro1_Heff3} H^{(3)}_{eff}&&=  \sum_{\langle ijk
\rangle}\Bigg((J_2 ({\bf S}_i \cdot {\bf e}^{(i)}_{ijk}) ({\bf
S}_j \cdot {\bf e}^{(j)}_{ijk})+
\\
\nonumber &&J_3 (\frac{1}{2}-{\bf S}_i \cdot {\bf e}^{(i)}_{ijk})
(\frac{1}{2}-{\bf S}_j \cdot {\bf e}^{(j)}_{ijk})
(\frac{1}{2}-{\bf S}_k \cdot {\bf e}^{(k)}_{ijk})\Bigg).
\end{eqnarray}

The coupling parameters (in units of $J$) are:  $ J_2=0.1049  $,  $ J_3=0.4215 
$,
  and $ {\bf e}^{(i)}_{123}, i=1,2,3$ are three unit vectors  in the x-y
plane
 whose angles $\a^{(i)}_{123}$ depend on the particular plane
defined by the triangle of tetrahedral units  $123$  as given in
table I of \cite{tsunetsugu}. The effective hamiltonian
(\ref{eq:pyro1_Heff3}) resembles the terms obtained by second
order perturbation theory (in inter-tetrahedra couplings) by
Harris \etal\cite{harris} and Tsunetugu\cite{tsunetsugu}. The
 classical mean field ground state of
(\ref{eq:pyro1_Heff3})  is identical to the
ground state found in Ref.\cite{tsunetsugu}: three of the four FCC
sublattices are ordered in the directions ${\bf e}(0),{\bf
e}(2\pi/3),{\bf e}(-2\pi/3)$, while the direction of the fourth is
completely degenerate.  Therefore,  classical mean field approximation for
(\ref{eq:pyro1_Heff3}) is insufficient to remove the
ground state degeneracy.  Tsunetsugu\cite{tsunetsugu} was able to
lift the degeneracy by including  spinwave fluctuations effects  which
produce  ordering at a new low energy scale.

\begin{figure}[h]
  \centering
  \includegraphics[width=6cm]{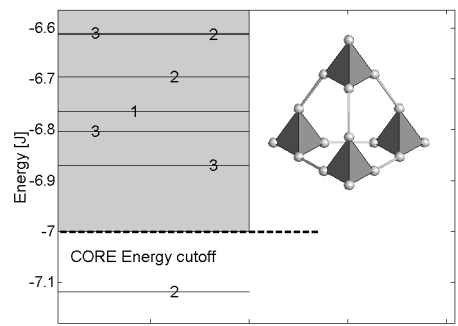}
  \caption{Low energy spectrum of a single ''supertetrahedron''.
  A dashed line illustrates the energy cutoff: only the
  two degenerate ground states are retained in the Hilbert space.}
  \label{fig:stspec}
\end{figure}

Here we  avoid the {\em a-priori}   symmetry breaking needed for
semiclassical spinwave theory, by treating  (\ref{eq:pyro1_Heff3})   fully 
quantum mechanically.  This entails a second CORE  transformation which
involves choosing  the  {\em
``supertetrahedron''}, as a basic cluster of four  tetrahedra,  whose structure
and spectrum are  depicted in  Fig.\ref{fig:stspec}. 

 Our new
 pseudospins ${\bf \tau}_i$
are defined by the two degerenate singlet ground states of the
supertetrahedron. (This degeneracy is found for  the
  Heisenberg
model on the original lattice as well as for  the effective 
 model
(\ref{eq:pyro1_Heff3})).
  These states transform as the E irreducible
representation of the tetrahedron ($T_d$)
  symmetry group, similarly to
the singlet ground states of a
 single tetrahedron.

The supertetrahedra form a cubic lattice, shown in Fig.
\ref{fig:pyrogs}. The effective hamiltonian (\ref{eq:pyro1_Heff3})
and the lattice geometry imply that  non-trivial effective
interactions appear only at the range of three supertetrahedra and higher.
Range three effective interactions include  two and three
pseudospin interactions, which are dominated by

\begin{eqnarray}
\label{eq:pyro2_Heff2} \mathcal H_{eff}&=&J_1 \sum_{\langle ij
\rangle}{({\bf \tau}_i \cdot {\bf f}_{ij}) ({\bf \tau}_j \cdot
{\bf f}_{ij})}+
\\
\nonumber && J^{(a)}_2 \sum_{\langle \langle ij \rangle
\rangle}{({\bf \tau}_i \cdot {\bf f}_{ij}) ({\bf \tau}_j \cdot
{\bf f}_{ij})}+
\\
\nonumber && J^{(b)}_2 \sum_{\langle  \langle ij \rangle
\rangle}{({\bf \tau}_i \cdot ({\bf f}_{ij} \times \hat {\bf z}))
({\bf \tau}_j \cdot ({\bf f}_{ij}\times \hat {\bf z})}).
\end{eqnarray}
Here,  $\langle ~\rangle$ and $\langle \langle~ \rangle \rangle$
indicate summation over nearest- and next
nearest-neighbors, respectively. The coupling constants  are found to be
relatively small:  $J_1=0.048J$,  $J^{(a)}_2=-0.006 J$  and
$J^{(b)}_2=0.018J$ . The vectors ${\bf f}_{ij}$
depend on the vector ${\bf r}_{ij}$ connecting the two sites, and
their values are presented in table \ref{tab:f_ij}.

\begin{table}[h]
\label{tab:f_ij} \caption{The values of the vectors ${\bf
f}_{ij}$ in eq. (\ref{eq:pyro2_Heff2}), depending on the vector
${\bf r}_{ij}$ separating the sites $i$ and $j$.}
\begin{tabular}{cc}
 ${\bf r}_{ij}$&${\bf f}_{ij}$\\ \hline \hline
 \begin{tabular}{c} {$(\pm 1,0,0)$}\\{$(0,\pm 1,\pm 1)$} \end{tabular}&$(1,0,0)$\\ \hline
 \begin{tabular}{c}{$(0,\pm 1,0)$}\\{$(\pm 1,0,\pm 1)$} \end{tabular}&$(-\frac{1}{2},\frac{\sqrt{3}}{2},0)$\\ \hline
 \begin{tabular}{c}{$(0,0,\pm 1)$}\\{$(\pm 1,\pm 1,0)$} \end{tabular}&$(-\frac{1}{2},-\frac{\sqrt{3}}{2},0)$\\

 \hline
\end{tabular}
\end{table}

We performed classical Monte Carlo simulations using the classical (large
spin) approximation to    (\ref{eq:pyro2_Heff2}). The  
ground state
 was found to choose an antiferromagnetic axis, and to be
ferromagnetic in the planes  as depicted in Fig. \ref{fig:pyrogs}.
This ground state differs from  the
semiclassical ground state\cite{harris,tsunetsugu}. The latter involves
condensation   of  high energy  states of the supertetrahedron in the
thermodynamic ground state.  Since in the  supertetrahedra 
diagonalizations, we find  a much  larger   energy gap than inter-site
coupling  we believe these excitations cannot condense to yield the
semiclassical ground state symmetry breaking.

 {\em Discussion}.  The CORE technique enabled  us to derive
effective Hamiltonian for highly frustrated antiferromagnets,
 written in
terms of low energy, local degrees of freedom. For
 both  Checkerboard and
Pyrochlore systems, we found lattice
 symmetry breaking ground states which are
essentially products of
 local singlets.  The spin gap to the lowest triplet 
excitation
 is large (of order $J$), and seems to survive  interplaquette
interactions. The low energy excitations are singlets, which are
local pseudospin flips, or Ising  domain walls between ground
states. The ordering  energy scale is of order $J/100$ for the
Pyrochlore.

This picture seems to
be consistent with existing numerical data for the density of low energy
singlets on the Checkerboard and Pyrochlore. Experimentally,   lattice
symmetry breaking could drive a  static  lattice distortion, which would
be  observable by
 additional  Bragg peaks in neutron and X-ray scattering.
For example, the antiferromagnetic
 order between planes of supertetrahedra
would correspond to a lattice
 distortion with wavelength of four tetrahedra.

{\em How general are these results?} 
Formation of local singlets is a natural way to relieve the frustration in
quantum antiferromagnets
that can be written as a sum over clusters $\sum_c(\sum_i{\bf S}_{ic})^2$.
On each even cluster the ground state is a singlet with a gap of order $1/S$
to a local triplet.
Frustration suppresses hopping of these triplets and could inhibit their condensation
into a spin ordered ground state.  
Thus lattice symmetry breaking singlet ground states are expected as 
a typical feature of frustrated quantum antiferromagnets\cite{read-sachdev}. 

{\em Acknowledgements.} We thank   A. Keren
S. Sondhi and O. Tchernyshyov for useful conversations. Support of
U.S.-Israel Binational Science Foundation, and the Fund for Promotion of
Research at Technion are gratefully acknowledged.


\begin{thebibliography}{10}
\bibitem{comm-MG} A similar  argument can  be applied  to prove the
dimerized  ground state for the Majumdar-Ghosh Hamiltonian.
\end{thebibliography}
 \end{document}